\documentstyle[agupp]{article}

\lefthead{SORRISO--VALVO ET AL.}
\righthead{PDF OF MHD FLUCTUATIONS}
\received{November 17, 1998}
\revised{January 29, 1999}
\accepted{Accepted March 16, 1999}

\authoraddr{R. Bruno and G. Consolini
Istituto di Fisica dello Spazio Interplanetario -- CNR, 
00133 Roma, Italy. 
(e-mail: bruno@ifsi.rm.cnr.it; consolini@ifsi.rm.cnr.it)}

\authoraddr{V. Carbone, L. Sorriso--Valvo and P. Veltri, 
Dipartimento di Fisica, Universit\'a della Calabria, 
87036 Roges di Rende (CS), Italy.
(e-mail: carbone@fis.unical.it; veltri2@fis.unical.it; 
veltri@fis.unical.it)}

\input epsf

\begin{document}

\title{Intermittency in the solar wind turbulence through probability 
distribution functions of fluctuations}

\author{Luca Sorriso--Valvo, Vincenzo Carbone and Pierluigi Veltri}
\affil{Dipartimento di Fisica, Universit\'a della Calabria 
       and Istituto Nazionale \\ per la Fisica della Materia, 
       Unit\'a di Cosenza, Italy}

\author{Giuseppe Consolini and Roberto Bruno}
\affil{Istituto di Fisica dello Spazio Interplanetario -- CNR, 00133 Roma, Italy}

\begin{abstract}

Intermittency in fluid turbulence can be emphasized through the analysis 
of Probability Distribution Functions (PDF) for velocity fluctuations, 
which display a strong non--gaussian behavior at small scales. Castaing et 
al. (1990) have introduced the idea that this behavior can be represented, 
in the framework of a multiplicative cascade model, by a convolution of 
gaussians whose variances is distributed according to a log--normal 
distribution. In this letter we have tried to test this conjecture on the 
MHD solar wind turbulence by performing a fit of the PDF of the bulk speed 
and magnetic field intensity fluctuations calculated in the solar 
wind, with the model. This fit allows us to calculate a parameter $\lambda^2$ 
depending on the scale, which represents the width of the log--normal 
distribution of the variances of the gaussians. The physical implications 
of the obtained values of the parameter as well as of its scaling law are 
finally discussed.

\end{abstract}

\section{Introduction}

The statistics of turbulent fluid flows can be characterized by the Probability 
Distribution Function (PDF) of velocity differences over varying scales 
(\markcite{Frisch}, 1995, and references therein). At large scales the PDF is 
approximately Gaussian, as the scale decreases, the wings of the distribution 
become increasingly stretched, so that large deviations from the average value 
are present. This phenomenon, usually ascribed to intermittency, has been 
observed and deeply 
investigated in fluid flows (\markcite{Frisch}, 1995, and references therein), 
and recently also in Magnetohydrodynamic (MHD) flows (see for example 
\markcite{Biskamp}, 1993; \markcite{Marsch and Tu}, 1997). Intermittency in 
MHD flows has been analyzed mainly by using satellite 
measurements of solar wind fluctuations (\markcite{Burlaga}, 1991; 
\markcite{Marsch and Liu}, 1993; \markcite{Carbone et al.}, 1995, 1996; 
\markcite{Ruzmaikin et al.}, 1995; \markcite{Horbury et al.}, 1997), or by 
using high resolution 2D numerical simulations 
(\markcite{Politano et al.}, 1998) and Shell Models (\markcite{Biskamp}, 1993, 
\markcite{Carbone}, 1994). All these analysis deal with the scaling 
exponents of structure functions, aimed to show that they follow anomalous 
scaling laws which can be compared with the usual energy cascade models for 
turbulence. 

The non gaussian nature of PDF in MHD solar wind turbulence has been 
evidentiated by \markcite{Marsch and Tu} (1994). In order to investigate 
the properties of intermittency through the analysis of non gaussian 
character of PDF, it would be necessary to quantify the departure of PDF 
from gaussian statistics and to analyze how this departure depends on the 
scale. Because of the idea of self--similarity underlying the energy cascade 
process in turbulence, Castaing and co--workers (\markcite{Castaing et al.}, 
1990) introduced a model which tries to characterize the behavior of the PDF's 
through the scaling law of a parameter describing how the shape of the PDF 
changes in going towards small scales (\markcite{Vassilicos}, 1995). In its 
simpler form the model can be 
introduced by saying that the PDF of the increments $\delta \psi$ 
(representing here both velocity and magnetic fluctuations) at a given scale 
$\tau$, is made by a convolution of the typical Gaussian distribution 
$P_G$, with a function $G_{\tau}(\sigma)$ which represents the 
weight of the gaussian distribution characterized by the variance $\sigma$

\begin{equation}
P_{\tau}\left(\delta \psi\right) = 
\int G_{\tau} \left(\sigma \right) 
P_G\left(\delta \psi, \sigma \right) d \sigma
\label{equ1}
\end{equation}
In the usual approach where the energy cascade 
is introduced through a fragmentation process, $\sigma$ is directly related 
to the local energy transfer rate $\epsilon$. In a self--similar situation, 
where the energy cascade generates only a scaling variation of $\sigma = 
<\delta \psi^2>^{1/2}$ according to the classical Kolmogorov's picture 
(\markcite{Frisch}, 1995), $G_{\tau}(\sigma)$ reduces to a Dirac function 
$G_{\tau}(\sigma) = \delta (\sigma-\sigma_0)$. In this case from eq. 
(\ref{equ1}) a Gaussian distribution $P_{\tau}(\delta \psi) = 
P_G(\delta \psi, \sigma_0)$ is recast. 
On the contrary when the cascade is not strictly self--similar, the width of 
the distribution $G_{\tau}$ is different from zero. In this way the scaling 
behavior of the width (which takes into account the height of the PDF's wings) 
can be used to characterize intermittency. In the present paper we will try 
to see if the departure from the gaussian statistics can be described within 
the framework of the cascade model (\ref{equ1}).

\section{Solar Wind Observations}

The satellite observations of both velocity and magnetic field in the 
interplanetary space, offer us an almost unique possibility to gain 
information on the turbulent MHD state in a very large scale range, say from 1 
AU (Astronomical Units) up to $10^3$ km. Since the aim of this letter is 
essentially to show that the PDF of solar wind fluctuations can be 
represented by the model (\ref{equ1}), we limit to analyse only plasma 
measurements of the bulk velocity $V(t)$ and magnetic field intensity $B(t)$. 
The detailed analysis of single velocity and magnetic field components 
fluctuations is left for a more extended work. 

We based our analysis on plasma measurements as recorded by the instruments on 
board Helios 2 during its 
primary mission in the inner heliosphere. The analysis period refers to the 
first 4 months of 1976 when the spacecraft orbited from $1$ AU, on day 17, to 
$0.29$ AU on day 108. The original data were collected in $81$ s bins and we 
choose a set of subintervals of 2 days each. The 
subintervals were selected separately within low speed regions and high speed 
regions. Fast wind was chosen having care of selecting a two--day interval 
within the trailing edge of each high speed stream. The choice was such 
that the average value of the wind speed was never below $550$ km/sec for 
all the "fast" intervals. Slow wind was selected picking up two--day 
intervals just before the stream--stream interface having care that the 
average speed value was never above $450$ km/sec for each interval. 
For each subinterval we calculated the velocity and magnetic 
increments at a given scale $\tau$ through $\delta V_{\tau} = V(t+\tau)-V(t)$ 
and $\delta B_{\tau} = B(t+\tau)-B(t)$, which represent characteristic 
fluctuations across eddies at the scale $\tau$. Then we normalize each 
variable to the standard deviation within each subinterval $\delta v_{\tau} = 
\delta V_{\tau}/[<(\delta V_{\tau})^2>]^{1/2}$ and $\delta b_{\tau} = 
\delta B_{\tau}/[<(\delta B_{\tau})^2>]^{1/2}$ 
(brackets being average within each subinterval at the scale $\tau$). Then we 
get two data sets: a set containing both the 
normalized velocity and magnetic fluctuations for the low speed streams (each 
variable is made by $10890$ samples), and a different set containing the same 
quantities for the high speed streams (each variable made of $13068$ samples). 
We calculate the PDF's at 11 different scales logarithmically spaced 
$\tau = \Delta t \; 2^n$, where $n=0,1,...,10$ and $\Delta t = 81$ s. We 
collect the number of events within each bins by using $31$ bins equally 
spaced in the range within $3$ times the standard deviation of the total 
sample. Before we mixed the different subperiods belonging to a given class 
(high or low speed streams), we tested for the fact that the gross features of 
PDF's shape does not change in different subintervals. Then our results for 
high and low speed streams are representative of what happens at the PDF's.

The results are shown in figures \ref{fig1} and \ref{fig2}, where we 
report the PDF's of both velocity and magnetic intensity for the high speed 
streams (the same figures can be done for the slow speed streams). At large 
scales the PDF's are almost Gaussian, and the wings of the distributions grow 
up as the scale becomes smaller. This is true in all cases, say for both 
types of wind. Stronger events at small scales have a probability of 
occurrence greater than that they would have if they were distributed 
according to a gaussian function. This behavior is at the heart of the 
phenomenon of intermittency as currently observed in fluid flows 
(\markcite{Frisch}, 1995) and in the solar wind turbulence 
(\markcite{Marsch and Tu}, 1997). As a characteristic it is worthwile to 
note that for the magnetic intensity, the PDF's wings at small scales are more 
"stretched" with respect to the corresponding PDF's calculated for velocity. 
This is true both in slow and fast wind. 

\section{Results and Discussion}

In order to make a quantitative analysis of the energy cascade leading to 
the process described in the previous section, we have tried to fit the 
distributions by using the log--normal ansatz 
(\markcite{Castaing et al.}, 1990).

\begin{equation}
G_{\tau} \left(\sigma \right) d\sigma = {1 \over \lambda(\tau) 
\sqrt{2\pi} } \exp \left[- {\ln^2 (\sigma / \sigma_0) \over 
2 \lambda^2(\tau)} \right] d (\ln \sigma)
\label{lognormal}
\end{equation}
even if also other functions gives rise to results not really different. The 
parameter $\sigma_0$ represents the most probable value of $\sigma$, while 
$\lambda(\tau) = 
<(\Delta \ln \sigma)^2>^{1/2}$ is the width of the log--normal distribution 
of $\sigma$.

\begin{figure}[h]
\epsfxsize=9cm    
\centerline{\epsffile{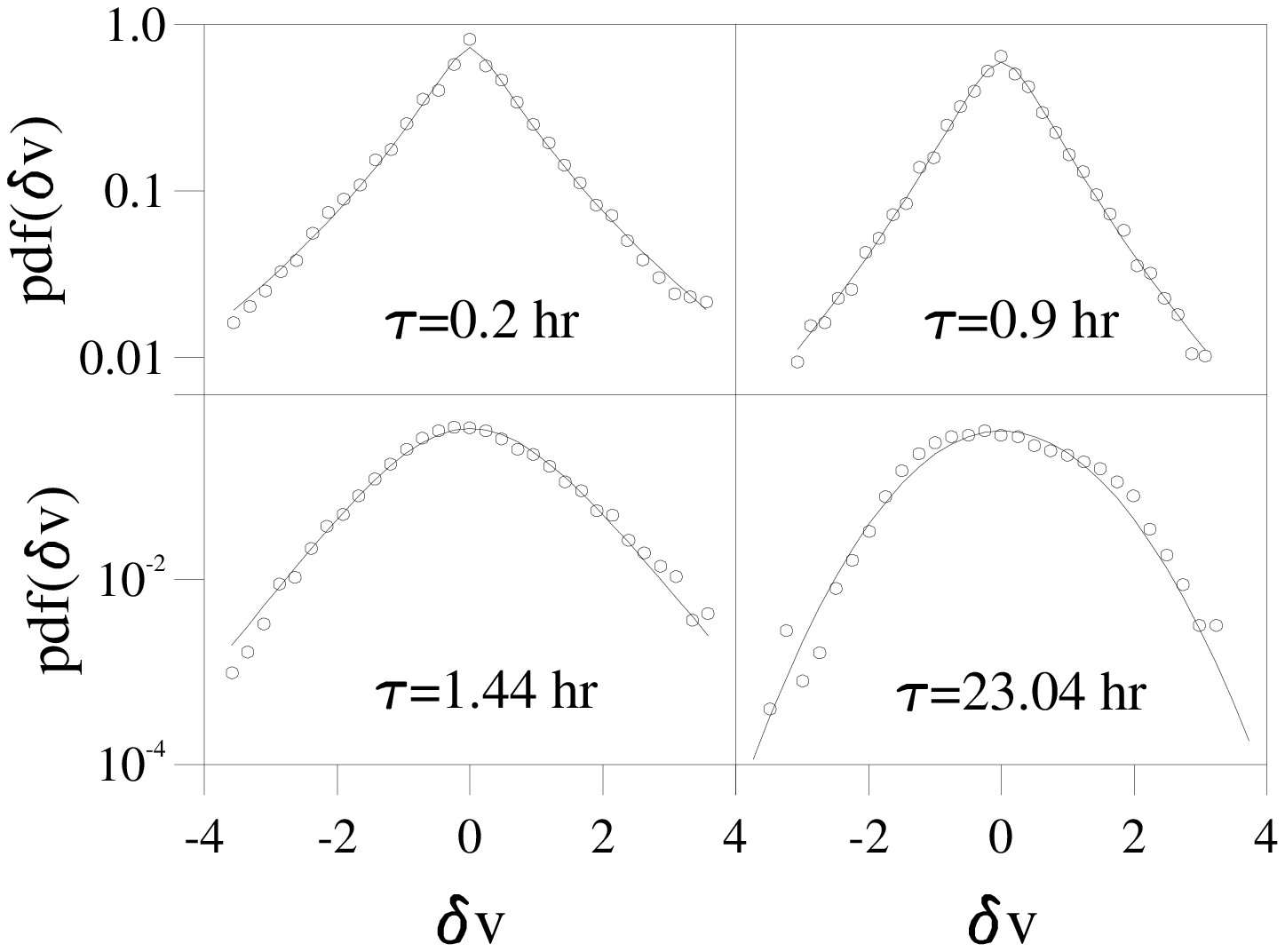}}  
\figurewidth{8cm}
\caption{The scaling behavior of the PDF for $\delta v_{\tau}$ as calculated 
from the experimental data (white symbols) in the fast streams. The full lines 
represent the fit obtained through the model as described in the text.}
\label{fig1}
\end{figure}

\begin{figure}[h]
\epsfxsize=9cm    
\centerline{\epsffile{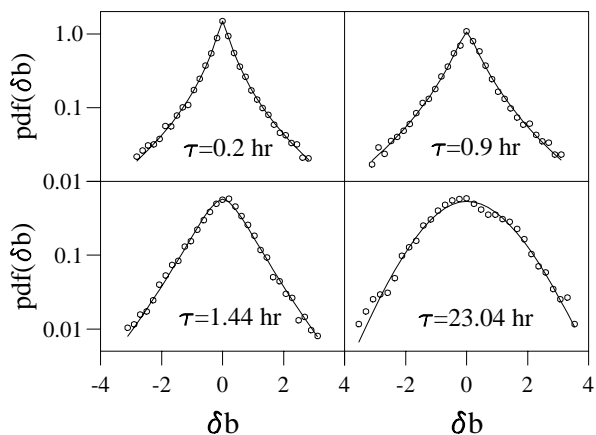}}  
\figurewidth{8cm}
\caption{The scaling behavior of the PDF for $\delta b_{\tau}$ as calculated 
from the experimental data (white symbols) in the fast streams. The full lines 
represent the fit obtained through the model as described in the text.}
\label{fig2}
\end{figure}

\begin{figure}[h]
\epsfxsize=7cm    
\centerline{\epsffile{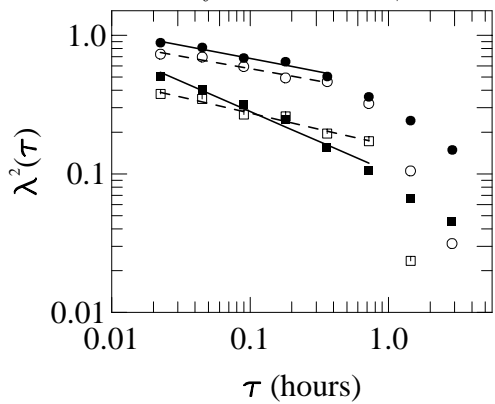}}  
\figurewidth{8cm}
\caption{We show the scaling behavior of $\lambda^2(\tau)$ vs. $\tau$ for both 
fast (black symbols) and slow (open symbols) streams. Circles refer to 
the magnetic field intensity, squares refer to the bulk velocity.}
\label{fig3}
\end{figure}

We have fitted the expression (\ref{equ1}) on the experimental PDF's for 
both velocity and magnetic intensity, and we have obtained the corresponding 
values of the parameter $\lambda$. The values of the parameters $\sigma_0$, 
which do not display almost any variation with $\tau$ are reported in the 
Table. Our results are summarized in figures 
\ref{fig1} and \ref{fig2}, where we plot, as full lines, the curves relative 
to the fit. As can be seen the scaling behavior of PDF's in all cases is very 
well described by the function (\ref{equ1}), thus indicating the robustness of 
the cascade model. From the fit, at each scale $\tau$, we get a value for 
the parameter $\lambda^2(\tau)$, and in figures \ref{fig3} we report the 
scaling behavior of $\lambda^2(\tau)$ for both high and low speed streams. 
Starting from $\lambda^2 \simeq 10^{-3}$ at the large scales (about $1$ day), 
the parameter increases abruptly to $\lambda^2 \simeq 10^{-1}$ at about $2$ 
hours, and finally a scaling law starts to become evident up to $\Delta t = 
81$ sec. In this last range, which corresponds roughly to what is usually 
called the "Alfv\'enic range", we fitted the parameter with a power law 
$\lambda^2(\tau) = \mu \tau^{-\beta}$. The values of $\mu$ and $\beta$ 
obtained in the fitting procedure and the corresponding range of scales, are 
reported in the Table. 

\begin{center}
\begin{planotable}{ccccc}
\tablewidth{40pc}
\tablecaption{We report the values of the parameters $\sigma_0$, and the 
values of $\mu$ and $\beta$ obtained in the fitting procedure for 
$\lambda^2(\tau)$. We also report the range of scales where the fit has been 
done.}
\tablenum{1}
\tablehead{\multicolumn{1}{c}{ } & 
\multicolumn{1}{c}{B (Fast)} & 
\multicolumn{1}{c}{B (Slow)} & 
\multicolumn{1}{c}{V (Fast)} & 
\multicolumn{1}{c}{V (Slow)}}
\startdata 
$\sigma_0$ & $0.85 \pm 0.05$ & $0.90 \pm 0.05$ & $0.90 \pm 0.05$ & 
             $0.95 \pm 0.05$ \nl
$\mu$ & $0.90 \pm 0.03$ & $0.75 \pm 0.03$ & $0.54 \pm 0.03$ & 
        $0.38 \pm 0.02$ \nl
$\beta$ & $0.19 \pm 0.02$ & $0.18 \pm 0.03$ & $0.44 \pm 0.05$ & 
          $0.20 \pm 0.04$ \nl
Scales & $\tau \leq 0.72$ hours & $\tau \leq 0.72$ hours & 
$\tau \leq 1.44$ hours & $\tau \leq 1.44$ hours  
\end{planotable}
\end{center}

Looking at Figure \ref{fig3}, it can be seen that both in fast and in slow 
streams magnetic field intensity is more intermittent than bulk speed (values 
of $\lambda^2$ are at least two times larger for magnetic field intensity than 
for velocity). This has also been reported by \markcite{Marsch and Tu} (1994), 
and the same indications comes from 2D MHD direct simulations 
(\markcite{Politano et al.}, 1998), and in analysis of solar wind 
intermittency performed using different thecniques (\markcite{Veltri and 
Mangeney}, 1999). The values of $\lambda^2(\tau)$ are more or less the same 
for magnetic field intensity both in fast and in slow wind. This is perhaps 
related to the fact that magnetic field intensity fluctuations are related 
to compressive fluctuations, which should have the same nature in both 
types of wind. The bulk velocity fluctuations on the contrary are more 
intermittent at small scales ($81$ sec) in the fast wind and at large scale 
($\simeq 1$ hour) in the slow wind. This result is due to the different 
values of $\beta$ for fast and slow wind. From the Table it appears that 
the value of $\beta$ is not universal, a result which has also been found 
in fluid flows (\markcite{Castaing et al.}, 1990) being close to $\beta \simeq 
0.2$ for magnetic field intensity in both fast and slow wind and for the 
velocity field in slow wind, while in fast wind the value of $\beta$ for 
the bulk velocity fluctuations is $\beta \simeq 0.44$.

In the framework of the cascade model, \markcite{Castaing et al.} (1990) give 
an interpretation of the parameter $\beta$ as the co--dimension of 
the more intermittent structures in a 1D cut of the turbulent field. If one 
believes to this interpretation, our results show that 
singular structures which are responsible for intermittency of the bulk 
velocity, look different for both type of winds. In particular structures in 
fast wind appears to lie on set with higher co--dimension. The fact that 
the value of $\beta$ for the bulk velocity fluctuations in slow wind is the 
same as the value of $\beta$ for magnetic field intensity suggests that 
intermittent structures in slow wind are perhaps mainly associated with 
compressive fluctuations. On the contrary the different value of $\beta$ 
found in fast wind evidentiate a different nature of velocity fluctuations 
in fast wind, perhaps related to the fact that such fluctuations are mainly 
incompressible. This result is in agreement with what has been recently 
found by \markcite{Veltri and Mangeney} (1999); these authors found that in 
fast wind the more intermittent structures are tangential discontinuities 
with almost no variation in magnetic field intensity, while in slow wind 
the most intermittent structures are shock waves, which display the same 
behavior in bulk velocity and magnetic field intensity.

\acknowledgments{We are grateful to H. Rosenbauer and R. Schwenn for making 
the Helios plasma data available to us.}


\end{document}